\documentclass[a4paper,twocolumn]{article}

\usepackage{amssymb,amsfonts,amsmath,amstext,amsgen,amsopn,amsxtra,indentfirst,graphicx,color}

\begin{document}

\title{The approximately universal shapes of epidemic curves in the Susceptible-Exposed-Infectious-Recovered (SEIR) model} 
\date{}
\maketitle
\noindent
\author{Kevin Heng$^{1,2\dagger}$, Christian L. Althaus$^3$}\\
\author{\scriptsize $^{1}$ University of Bern, Center for Space and Habitability, Gesellschaftsstrasse 6, CH-3012, Bern, Switzerland. Email: kevin.heng@csh.unibe.ch}\\
\author{\scriptsize $^{2}$ University of Warwick, Department of Physics, Astronomy \& Astrophysics Group, Coventry CV4 7AL, United Kingdom. Email: Kevin.Heng@warwick.ac.uk}\\
\author{\scriptsize $^{3}$ University of Bern, Institute of Social and Preventive Medicine, Mittelstrasse 43, CH-3012, Bern, Switzerland. Email: christian.althaus@ispm.unibe.ch}\\
\author{\scriptsize $^{\dagger}$ Corresponding author}

\begin{abstract}
Compartmental transmission models have become an invaluable tool to study the dynamics of infectious diseases. The Susceptible-Infectious-Recovered (SIR) model is known to have an exact semi-analytical solution.  In the current study, the approach of Harko et al. (2014) is generalised to obtain an approximate semi-analytical solution of the Susceptible-Exposed-Infectious-Recovered (SEIR) model.  The SEIR model curves have nearly the same shapes as the SIR ones, but with a stretch factor applied to them across time that is related to the ratio of the incubation to infectious periods.  This finding implies an approximate characteristic timescale, scaled by this stretch factor, that is universal to all SEIR models, which only depends on the basic reproduction number and initial fraction of the population that is infectious. 
\end{abstract}

\section{Introduction}

Compartmental models provide a key tool in infectious disease epidemiology for studying the transmission dynamics of various pathogens \cite{vwbook,kn10,ws18}. The Susceptible-Infectious-Recovered (SIR) model is known to have an exact semi-analytical solution \cite{km1927,harko2014,miller2017}. No such solution exists for the Susceptible-Exposed-Infectious-Recovered (SEIR) model, although some of its properties have been examined using an approximate analytical approach \cite{pio20}. In the current study, the approach of \cite{harko2014} is generalised to demonstrate that, while no exact semi-analytical solution is possible, an approximate one does exist.

It will be demonstrated that this approximate solution of the SEIR model implies the curves of \textit{all} SEIR models are simply stretched or compressed relative to one another by the factor,
\begin{equation}
\alpha = \frac{\sigma}{\sigma + \gamma},
\end{equation}
where the incubation period is $1/\sigma$, the infectious period is $1/\gamma$ and the generation time is $1/\sigma+1/\gamma$. The SIR model is a special case with $\alpha=1$. This property implies the time taken for the infectious curve to peak is approximately universal for the SEIR model when scaled by $\alpha$.

In Section \ref{sect:SIR}, the SIR model is concisely reviewed and extended.  In Section \ref{sect:SEIR}, approximate solutions of the SEIR model and their implications are elucidated.  A concise summary is provided in Section \ref{sect:summary}.

\begin{figure*}[!h]
\begin{center}
\vspace{-0.2in}
\includegraphics[width=\columnwidth]{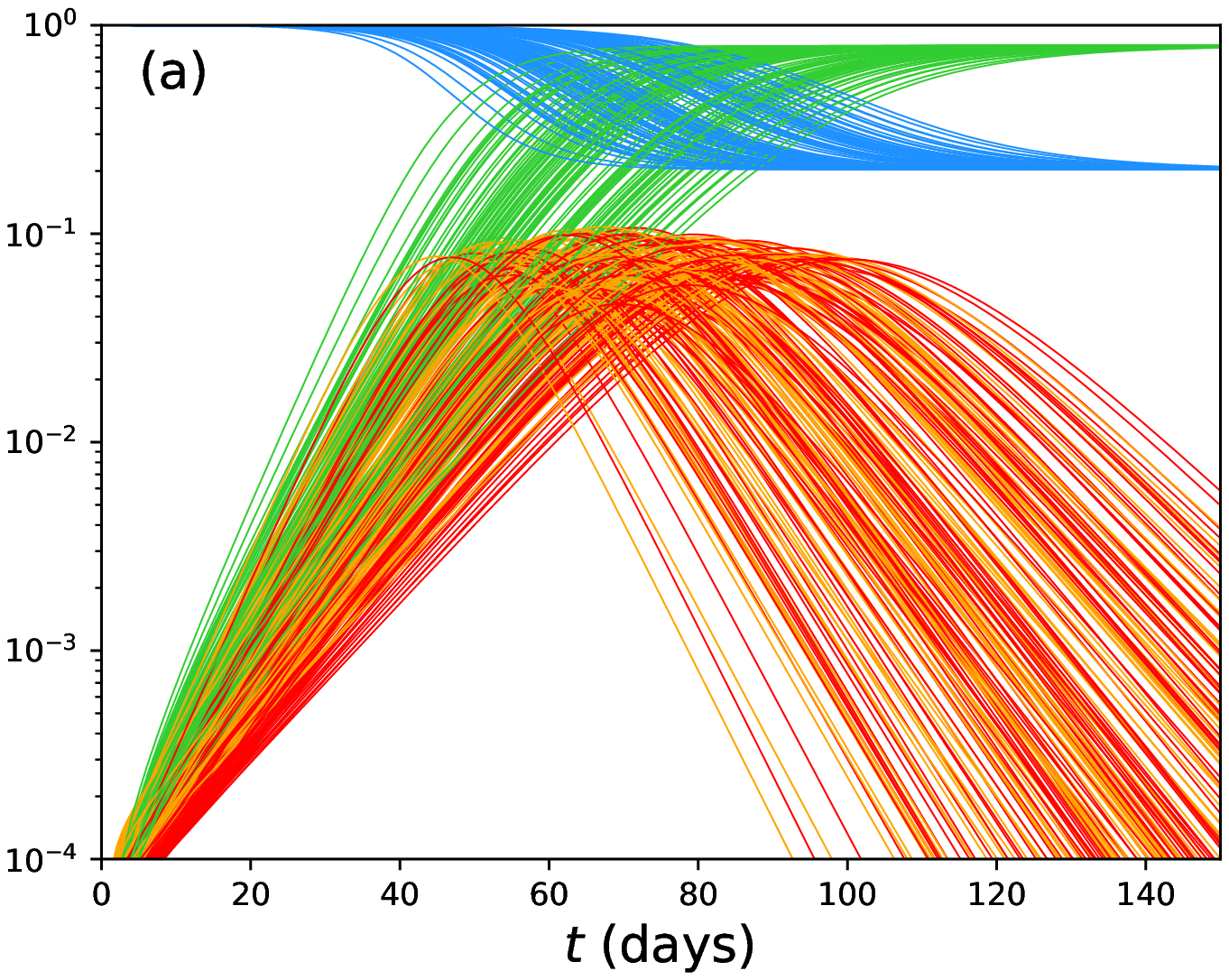}
\includegraphics[width=\columnwidth]{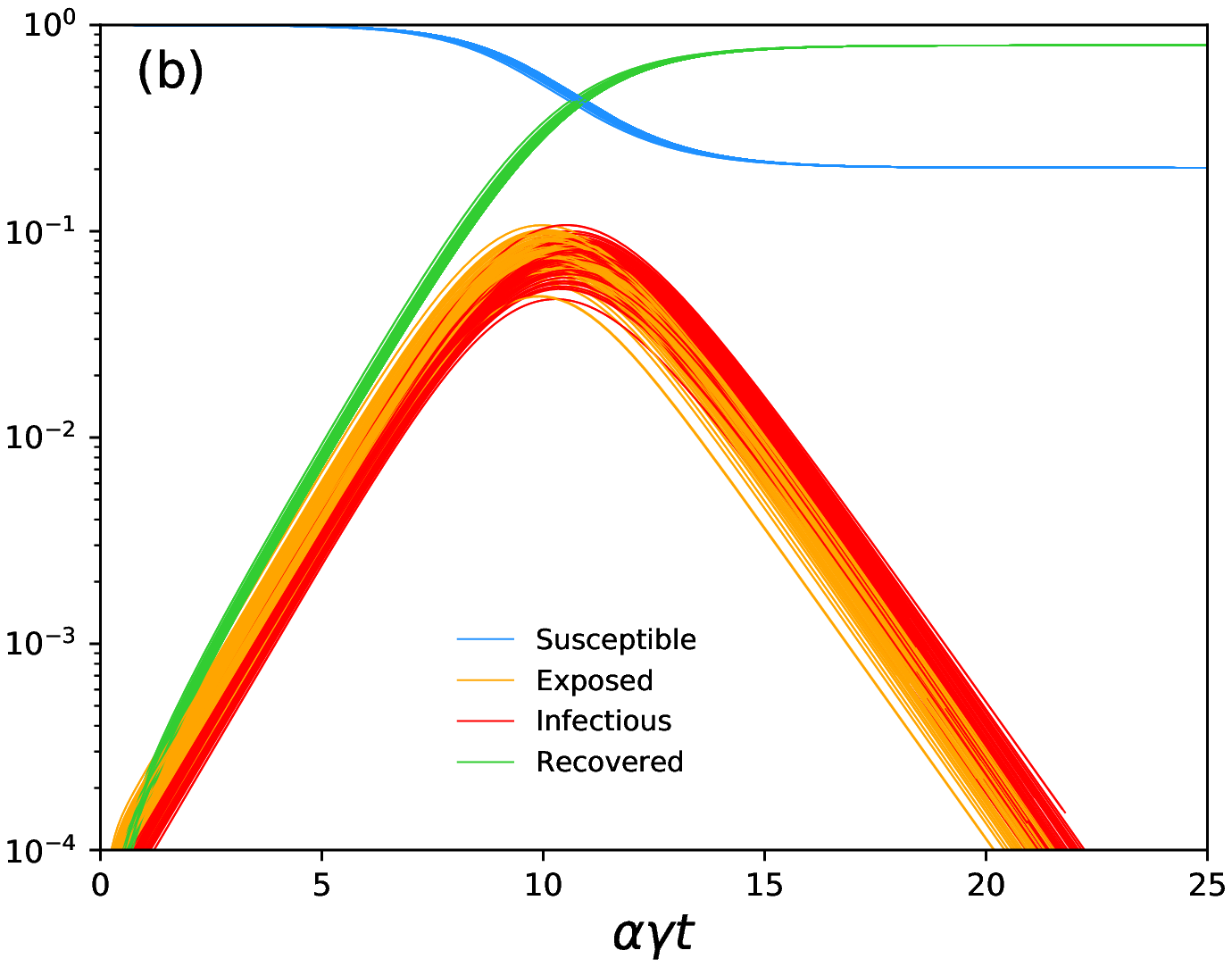}
\end{center}
\vspace{-0.2in}
\caption{\scriptsize Solution curves of 100 SEIR models as a (a) function of time and (b) time scaled by $\alpha \gamma$.  For illustration, the basic reproduction number has been set to ${\cal R}_0=2$ and the initial fraction of the population that is infectious has been set to $I_0=10^{-4}$.  Each set of curves is generated using 100 random realisations of the incubation and infectious periods, each drawn from an interval between 2 and 5 days for illustration.}
\vspace{-0.1in}
\label{fig:shapes}
\end{figure*}

\section{The SIR Model}
\label{sect:SIR}

In the SIR model, the fraction of the population that is susceptible ($S$) becomes infected at a rate $\beta = {\cal R}_0 \gamma$, where ${\cal R}_0$ is the basic reproduction number. There is no incubation period. The fraction of the population that is infected is immediately infectious ($I$) for a period of $1/\gamma$, after which a fraction of the population recovers ($R$). The SIR model is described by the following set of coupled ordinary differential equations \cite{vwbook,harko2014},
\begin{equation}
\begin{split}
\frac{dS}{dt} &= - \beta IS, \\
\frac{dI}{dt} &= \beta IS - \gamma I, \\
\frac{dR}{dt} &= \gamma I, \\
\end{split}
\label{eq:SIR}
\end{equation}
where $t$ represents the time. Since this set of equations does not consider births or deaths, we have $S+I+R=1$.

\subsection{Review of Harko et al. (2014)}
\label{subsect:harko14}

As a starting point, the derivation of \cite{harko2014} is made more compact and cast in the mathematical notation of the current study.  By taking the derivative of the first equation of (\ref{eq:SIR}) with respect to time, one obtains equation (12) of \cite{harko2014},
\begin{equation}
I^{\prime} = - \frac{1}{\beta} \left[ \frac{S^{\prime\prime}}{S} - \left( \frac{S^{\prime}}{S} \right)^2 \right],
\label{eq:harko12}
\end{equation}
where for convenience one uses shorthand notation for the derivatives with respect to time,
\begin{equation}
I^{\prime} \equiv \frac{dI}{dt}, ~S^{\prime} \equiv \frac{dS}{dt}, ~S^{\prime\prime} \equiv \frac{d^2S}{dt^2}.
\end{equation}
By combining equation (\ref{eq:harko12}) with the second equation in (\ref{eq:SIR}), one obtains equation (13) of \cite{harko2014},
\begin{equation}
\frac{S^{\prime\prime}}{S} - \left( \frac{S^{\prime}}{S} \right)^2 + \frac{\gamma S^\prime}{S} - \beta S^\prime = 0.
\label{eq:harko13}
\end{equation}

By using the change of variables,
\begin{equation}
S^{\prime} = \phi^{-1}, ~S^{\prime\prime} = -\phi^{-3} \frac{d\phi}{dS},
\end{equation}
one obtains from equation (\ref{eq:harko13}) an expression that is equivalent, but not identical, to equation (24) of \cite{harko2014},
\begin{equation}
\frac{d\phi}{dS} + \frac{\phi}{S} + \left( \beta S - \gamma \right) \phi^2 = 0,
\end{equation}
because one has chosen to work directly with $S$ (and not $S/S_0$) as the independent variable.  The preceding expression is recognised as a Bernoulli differential equation, which may be solved to obtain an expression that is equivalent, but not identical, to equation (25) of \cite{harko2014},
\begin{equation}
\phi^{-1} = S \left[ \beta \left(S - S_0 - I_0 \right) - \gamma \ln{\left(\frac{S}{S_0}\right)} \right],
\end{equation}
where the initial value of $S$ is denoted as $S_0$.  The constant of integration is set by demanding that $S+I+R=1$.  Recalling the definition of $\phi$, an expression that is equivalent to equation (26) of \cite{harko2014} follows,
\begin{equation}
t - t_0 = \int^S_{S_0} \frac{1}{s \left[ \beta \left(s - S_0 - I_0 \right) - \gamma \ln{\left(\frac{s}{S_0}\right)} \right]} ~ds,
\label{eq:harko_integral}
\end{equation}
where $t_0$ is the initial time.  The preceding integral has no exact analytical (closed-form) solution and needs to be evaluated numerically, which is why it is strictly speaking an exact semi-analytical solution of the SIR model.

The first and third equation of (\ref{eq:SIR}) may be combined to obtain
\begin{equation}
R = \frac{\gamma}{\beta} \ln{\left( \frac{S_0}{S} \right)},
\end{equation}
where the initial fraction of the population that has recovered is chosen to be $R_0=0$, which in turn implies that the initial fraction of the population that is infectious is $I_0 = 1 - S_0$.

\subsection{Extension of Harko et al. (2014)}

By setting $I^\prime=0$ in equation (\ref{eq:SIR}), one realizes that the infectious curve $I$ peaks at $S=\gamma/\beta = 1/{\cal R}_0$.  Thus, equation (\ref{eq:harko_integral}) may be used to express the time taken for $I$ to peak,
\begin{equation}
\gamma ~\Delta t \approx \int^{1/{\cal R}_0}_{S_0} \frac{1}{S \left[ {\cal R}_0 \left(S - S_0 \right) - \ln{\left(\frac{S}{S_0}\right)} \right]} ~dS,
\label{eq:timescale}
\end{equation}
where one assumes $I_0 \ll 1$.  The quantity $\gamma \Delta t$ is the time interval expressed in terms of the infectious period and depends only on two parameters: ${\cal R}_0$ and $I_0$.  Variations in $I_0$ shift the $S$, $I$ and $R$ curves back and forth in time without changing their shapes.  We emphasize a subtle choice of notation: $R_0$ is the initial fraction of the population that has recovered (and is always set to zero in the current study), whereas ${\cal R}_0$ is the basic reproduction number.

When the infectious curve $I$ first starts to rise from its initial value, the logarithm term in the integral of equation (\ref{eq:harko_integral}) may be approximated as $\ln{(S/S_0)} \approx S/S_0 - 1$, which allows the integral to be evaluated analytically. It follows that
\begin{equation}
\begin{split}
S &\approx \Lambda \left[ \gamma \left( {\cal R}_0 - \frac{1}{S_0} \right) + \frac{\gamma {\cal R}_0 I_0}{S_0} e^{\Lambda \left(t-t_0\right)} \right]^{-1}, \\
I &\approx 1 - \frac{1}{{\cal R}_0} - \left( 1 - \frac{1}{S_0 {\cal R}_0} \right) S,
\end{split}
\end{equation}
where we have defined the epidemic growth rate as
\begin{equation}
\Lambda \equiv \gamma \left( {\cal R}_0 - 1 \right),
\end{equation}
from which one obtains the known relationship between the basic reproduction number and the growth rate \cite{vwbook,Wallinga2007},
\begin{equation}
{\cal R}_0 = 1 + \Lambda D,
\label{eq:SIR_growth_rate}
\end{equation}
where $D \equiv 1/\gamma$ is the infectious period.  

\section{The SEIR Model}
\label{sect:SEIR}

\begin{figure}[!t]
\begin{center}
\vspace{-0.2in}
\includegraphics[width=\columnwidth]{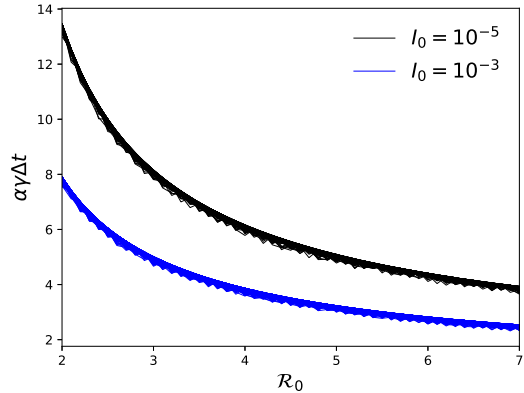}
\end{center}
\vspace{-0.2in}
\caption{\scriptsize Time until the infectious curve ($I$) peaks as a function of the basic reproduction number ${\cal R}_0$. In the SEIR model, the time to the epidemic peak ($\Delta t$) scales approximately with $\alpha$ and $\gamma$. For illustration, two values of the initial fraction of population that is infectious ($I_0$) are considered.  Each set of curves is generated using 10,000 random draws of the incubation and infectious periods from an interval between 2 and 5 days.}
\label{fig:timescale}
\end{figure}

\subsection{Seeking An Approximate Semi-Analytical Solution}

The SEIR model builds on the SIR model by considering an additional compartment for the fraction of the population that is exposed ($E$): infected but not yet infectious. The incubation period is $1/\sigma$. The SEIR model is described by the following set of coupled ordinary differential equations \cite{vwbook},
\begin{equation}
\begin{split}
\frac{dS}{dt} &= - \beta IS, \\
\frac{dE}{dt} &= \beta IS - \sigma E, \\
\frac{dI}{dt} &= \sigma E - \gamma I, \\
\frac{dR}{dt} &= \gamma I. \\
\end{split}
\label{eq:SEIR}
\end{equation}
Since this set of equations does not consider births or deaths, we have $S+E+I+R=1$.

The first and fourth equations may be combined to obtain
\begin{equation}
R = \frac{\gamma}{\beta} \ln{\left( \frac{S_0}{S} \right)},
\label{eq:R}
\end{equation}
which is identical to the SIR model.  Again, the choice of $R_0=0$ is made with no loss of generality.

By combining all four equations, one obtains
\begin{equation}
\frac{d^3R}{dt^3} + \left( \sigma + \gamma \right) \frac{d^2R}{dt^2} + \sigma \gamma \left( \frac{dR}{dt} + \frac{dS}{dt} \right) = 0.
\label{eq:R3}
\end{equation}
The approximation is taken that the rate of change of the acceleration of $R$ is vanishingly small,
\begin{equation}
R^{\prime\prime\prime} \equiv \frac{d^3R}{dt^3} = 0.  
\end{equation}
This yields
\begin{equation}
\frac{d^2R}{dt^2} + \alpha \gamma \left( \frac{dR}{dt} + \frac{dS}{dt} \right) = 0,
\end{equation}
where one defines $\alpha \equiv \sigma/(\sigma+\gamma)$.  When $\alpha=1$, one recovers equation (19) of \cite{harko2014} for the SIR model. 

One generalises equation (13) of \cite{harko2014},
\begin{equation}
\frac{S^{\prime\prime}}{S} - \left( \frac{S^{\prime}}{S} \right)^2 + \frac{\alpha \gamma S^\prime}{S} - \alpha \beta S^\prime = 0,
\end{equation}
from which the familiar Bernoulli equation follows,
\begin{equation}
\frac{d\phi}{dS} + \frac{\phi}{S} + \alpha \left( \beta S - \gamma \right) \phi^2 = 0.
\end{equation}
Retaining the $R^{\prime\prime\prime}$ term in equation (\ref{eq:R3}) would lead to a second-order, non-linear ordinary differential equation of $\phi(S)$ with no known analytical solution.

Solving for $\phi$ as in Section \ref{subsect:harko14} yields
\begin{equation}
\phi^{-1} = S \left[ \frac{1}{S_0 \phi_0} + \alpha \beta \left(S - S_0 \right) - \alpha \gamma \ln{\left(\frac{S}{S_0}\right)} \right],
\end{equation}
where $\phi_0$ is the initial value of $\phi$.  The preceding expression leads to an expression for $I$, in terms of $S$, with a yet unspecified constant of integration ($\phi_0$),
\begin{equation}
I = - \frac{1}{\beta S_0 \phi_0} - \alpha \left(S - S_0 \right) + \frac{\alpha \gamma}{\beta} \ln{\left(\frac{S}{S_0}\right)}.
\end{equation}

Let the initial fraction of the population that is exposed be $E_0$.  Demanding that $S_0 + E_0 + I_0 + R_0 = 1$ yields
\begin{equation}
I_0 = -\frac{1}{\beta S_0 \phi_0} = 1 - S_0 - E_0.
\end{equation}
Expressions for $E$ and $I$, in terms of $S$, are obtained
\begin{equation}
\begin{split}
E &= 1 - I_0 - \alpha S_0 + \left( \alpha - 1 \right) \left[ S - \frac{\gamma}{\beta} \ln{\left(\frac{S}{S_0}\right)} \right], \\
I &= I_0 - \alpha \left(S - S_0 \right) + \frac{\alpha \gamma}{\beta} \ln{\left(\frac{S}{S_0}\right)}.
\end{split}
\label{eq:E_and_I}
\end{equation}
Finally, $S$ can be expressed in terms of $t$ via the following integral,
\begin{equation}
t - t_0 = \int^S_{S_0} \frac{1}{s \left\{ \beta \left[ -I_0 + \alpha \left(s - S_0 \right) \right] - \alpha \gamma \ln{\left(\frac{s}{S_0}\right)} \right\}} ~ds.
\label{eq:Sintegral}
\end{equation}

Since $I_0 \ll 1$, the time taken for $I$ to peak is
\begin{equation}
\alpha \gamma ~\Delta t \approx \int^{1/{\cal R}_0}_{S_0} \frac{1}{S \left[ {\cal R}_0 \left(S - S_0 \right) - \ln{\left(\frac{S}{S_0}\right)} \right]} ~dS.
\label{eq:Sintegral2}
\end{equation}
The preceding expression is identical to equation (\ref{eq:timescale}) of the SIR model, except for the extra factor of $\alpha$.  It should be noted that the upper limit of the integral ($1/{\cal R}_0$) assumes the approximation $I^\prime=E^\prime=0$. However, equation (\ref{eq:Sintegral2}) is not used to compute the peak times in Figure \ref{fig:timescale}. Its only purpose is to demonstrate that one may factor out $\alpha \gamma$ from the integral. Stating the upper limit of the integral in equation (\ref{eq:Sintegral2}) more accurately does not alter the main conclusion of the current study.

The relationship between the growth rate and the basic reproduction number can again be derived. Using the same series expansion of the logarithm term in the integral of equation (\ref{eq:Sintegral}), one obtains
\begin{equation}
\begin{split}
S &\approx \Lambda \left[ \alpha \gamma \left( {\cal R}_0 - \frac{1}{S_0} \right) + \frac{\gamma {\cal R}_0 I_0}{S_0} e^{\Lambda \left(t-t_0\right)} \right]^{-1}, \\
I &\approx I_0 + \alpha \left( S_0 - \frac{1}{{\cal R}_0} \right) - \left( 1 - \frac{1}{S_0 {\cal R}_0} \right) \alpha S,
\end{split}
\end{equation}
albeit with a different definition of the growth rate,
\begin{equation}
\Lambda \equiv \gamma {\cal R}_0 \left( I_0 + \alpha S_0 \right) - \alpha \gamma.
\end{equation}
It follows that
\begin{equation}
{\cal R}_0 = \frac{\alpha + \Lambda D}{I_0 + \alpha S_0} = \frac{1 + \Lambda \left( D^\prime + D \right)}{S_0 + I_0 \left( 1 + \frac{D^\prime}{D} \right)},
\label{eq:SEIR_growth_rate}
\end{equation}
where $D^\prime \equiv 1/\sigma$ is the incubation period. When $\alpha=1$, the expression for the SIR model in equation (\ref{eq:SIR_growth_rate}) is recovered. If $S_0 \approx 1$ and $I_0 \ll 1$, then one obtains ${\cal R}_0 \approx 1 + \Lambda (D + D^\prime)$.

The exact relationship between the growth rate and ${\cal R}_0$ has been derived in various ways \cite{Wallinga2007} (and references therein) and is given by ${\cal R}_0 = (1 + \Lambda D^\prime)(1+\Lambda D)$. This equation accounts for the characteristic generation time distribution of SEIR models, which is a convolution of the exponentially distributed incubation and infectious periods with mean durations of $D^\prime$ and $D$, respectively. The approximate solution of equation (\ref{eq:SEIR_growth_rate}) lacks the term $\Lambda^2 D^\prime D$. Hence, it corresponds to the case of an exponentially distributed generation time with mean duration $D^\prime + D$, which is the same as the solution for the SIR model assuming an infectious period of $D^\prime + D$.

\subsection{Implications}

Equation (\ref{eq:Sintegral2}) has non-trivial implications.  It suggests that the susceptible, exposed, infectious and recovered curves of SEIR models, with different values of $D^\prime$ and $D$, follow approximately universal shapes that are stretched by a factor of $1/\alpha = 1 + D^\prime/D$ relative to one another.  To demonstrate this property, the full set of coupled equations in (\ref{eq:SEIR}) is solved numerically using the \texttt{solve\_ivp} routine of the Python programming language suite \cite{virtanen2020}.  For illustration, one assumes ${\cal R}_0=2$ and $I_0=10^{-4}$.   Figure \ref{fig:shapes} shows the solution curves of 100 SEIR models, where the values of the incubation ($D^\prime \equiv 1/\sigma$) and infectious ($D \equiv 1/\gamma$) periods are randomly drawn from an interval between 2 and 5 days.  When time is scaled by the factor $\alpha \gamma$, the 100 susceptible, exposed, infectious and recovered curves lie approximately on top of one another.  

The second implication is that the time taken for the infectious curve to peak is approximately universal for \textit{all} SEIR models when scaled by $\alpha$ and expressed in terms of the infectious period.  In other words, $\alpha \gamma \Delta t$ should only depend on ${\cal R}_0$ and $I_0$.  To demonstrate this property, the full set of equations in (\ref{eq:SEIR}) is again solved numerically for 10,000 random draws of $1/\sigma$ and $1/\gamma$ and for ${\cal R}_0=2$ to 7.  For each SEIR model, the time taken for the infectious curve to peak ($\Delta t$) is calculated numerically.  All 10,000 values of $\Delta t$ are multiplied by $\alpha \gamma$; two sets of curves with different $I_0$ values are shown in Figure \ref{fig:timescale} for illustration.  For all 10,000 SEIR models, the $\alpha \gamma \Delta t$ values lie approximately on the same curve across ${\cal R}_0$ for a given value of $I_0$, demonstrating that $\alpha \gamma \Delta t$ is a dimensionless (with no physical units), approximately universal timescale of the SEIR model.

\section{Summary}
\label{sect:summary}

In the current study, approximate semi-analytical solutions of the SEIR model are found by generalising a previous approach for deriving an exact solution of the SIR model.  This finding implies that the entire family of susceptible, exposed, infectious and recovered curves of the SEIR model follow approximately universal shapes that are stretched or compressed, relative to one another, by a factor consisting of the incubation and infectious periods. The time taken for the infectious curve to peak is the characteristic timescale of the system and depends only on the basic reproduction number and the initial fraction of the population that is infectious when scaled by the infectious period and this stretch factor.

\scriptsize

\vspace{0.4in}

\textit{K.H. formulated the problem, derived the equations, performed the numerical calculations and wrote the manuscript. C.L.A. made the link between the reproduction number and growth rate, checked the equations and edited the manuscript.}

\end{document}